**Space-like energy-momentum in quantum electrodynamics**

By

Dan Solomon

Rauland-Borg Corporation
3450 W. Oakton
Skokie, IL

Email: dan.solomon@rauland.com
Phone: 847-324-8337

March 25, 2007

**Abstract**

A common assumption in quantum field theory is that the energy-momentum 4-vector of any quantum state must be time-like. However it has been recently shown [4] that this is not the case for a Dirac-Maxwell field in the coulomb gauge. Here we will present a proof that is simpler then the proof of Ref. [4] that there must exist quantum states which are space-like for a Dirac-Maxwell field in the coulomb gauge.

**1. Introduction**

A common assumption in quantum field theory is that the energy-momentum 4-vector of quantum states must be time-like. (See, for example, page 58 of Haag[1], Section 8.1 of Nishijima[2], and Section 10.7 of Weinberg[3]). While this is certainly the case for non-interacting (free) fields it has been shown [4] that this is not the case for a Dirac-Maxwell field in the coulomb gauge. In this paper we will revisit this problem and provide an alternative and simpler proof that there must exist quantum states with space-like energy-momentum.

Let $\hat{H}$ be the Hamiltonian operator and $\hat{P}_j$ the momentum operator where $j = 1, 2, 3$. We make the usual assumption [2] that there exist a complete orthonormal set of eigenstates $\left| \varphi_n \right\rangle$ which satisfy,

$$\hat{H} \left| \varphi_n \right\rangle = E_n \left| \varphi_n \right\rangle \text{ where } E_n \geq 0 \,; \quad \hat{P}_j \left| \varphi_n \right\rangle = P_{j,n} \left| \varphi_n \right\rangle \tag{1.1}$$

and

$$\left\langle \varphi_n \middle| \varphi_m \right\rangle = \delta_{nm} \,; \qquad \sum_n \left| \varphi_n \right\rangle \left\langle \varphi_n \right| = 1 \tag{1.2}$$

It is generally assumed that the energy-momentum 4-vector of each eigenstate $\left| \varphi_n \right\rangle$ is time-like. This means that the quantity $G_n = E_n^2 - \sum_{j=1}^{3} P_{j,n}^2 \geq 0$ for all $\left| \varphi_n \right\rangle$. However it will be shown that this is assumption is not correct for a Dirac-Maxwell field in the coulomb gauge and that there must exist eigenstates for which $G_n < 0$. In this case the energy-momentum is said to be space-like.

Define the operator,

$$\hat{F} = \hat{H} - \hat{P}_1 \tag{1.3}$$

From the above discussion we obtain,

$$\hat{F} \left| \varphi_n \right\rangle = F_n \left| \varphi_n \right\rangle \text{ where } F_n = E_n - P_{1,n} \tag{1.4}$$

If $F_n < 0$ then it is evident that $G_n < 0$. It will be shown in the following discussion that there must exist at least one eigenstate for which $F_n < 0$ so that energy-momentum 4-vector of this eigenstate is space-like.

## 2. A Dirac-Maxwell system

In the coulomb gauge the field operators for a Dirac-Maxwell system obey Maxwell's equations,

$$i\left[\hat{H}, \hat{\mathbf{E}}\right] = \nabla \times \hat{\mathbf{B}} - \hat{\mathbf{j}}; \quad i\left[\hat{H}, \hat{\mathbf{B}}\right] = -\nabla \times \hat{\mathbf{E}} \qquad (2.1)$$

where $\hat{\mathbf{E}}(\mathbf{x})$ is the electric field operator, $\hat{\mathbf{B}}(\mathbf{x})$ is the magnetic field operator, $\hat{\mathbf{j}} = \hat{\psi}^\dagger \boldsymbol{\alpha} \hat{\psi}$ is the current operator, and $\hat{\psi}(\mathbf{x})$ and $\hat{\psi}^\dagger(\mathbf{x})$ are the fermion field operators. The commutator between a local field operator $\hat{O}(\mathbf{x})$ and the momentum operator $\hat{P}_j$ is [2,5],

$$i\left[P_j, \hat{O}(\mathbf{x})\right] = -\frac{\partial \hat{O}(\mathbf{x})}{\partial x_j} \qquad (2.2)$$

Use this to obtain,

$$i\left[P_j, \hat{\mathbf{E}}(\mathbf{x})\right] = -\frac{\partial \hat{\mathbf{E}}(\mathbf{x})}{\partial x_j}; \qquad i\left[P_j, \hat{\mathbf{B}}(\mathbf{x})\right] = -\frac{\partial \hat{\mathbf{B}}(\mathbf{x})}{\partial x_j} \qquad (2.3)$$

Define the operator,

$$\hat{U} = e^{-i\hat{C}} \qquad (2.4)$$

where,

$$\hat{C} = \int\left[\left(\mathbf{D} \cdot \nabla \times \hat{\mathbf{E}}\right) + \mathbf{L} \cdot \hat{\mathbf{B}}\right] d^3 x \qquad (2.5)$$

where $\mathbf{D}(\mathbf{x})$ and $\mathbf{L}(\mathbf{x})$ are real vector functions that will be specified later. Now $\hat{\mathbf{E}} = \hat{\mathbf{E}}_T + \hat{\mathbf{E}}_L$ where $\hat{\mathbf{E}}_T$ is the transverse part of the electric field and $\hat{\mathbf{E}}_L$ is the longitudinal part. From the definition of the curl we have $\nabla \times \hat{\mathbf{E}} = \nabla \times \hat{\mathbf{E}}_T$. Use this in (2.5) to obtain,

$$\hat{C} = \int\left[\left(\mathbf{D} \cdot \nabla \times \hat{\mathbf{E}}_T\right) + \mathbf{L} \cdot \hat{\mathbf{B}}\right] d^3 x \qquad (2.6)$$

Due to the fact that $\hat{\mathbf{E}}_T$ and $\hat{\mathbf{B}}$ are real we have that $\hat{C}^\dagger = \hat{C}$. Therefore $\hat{U}^\dagger = e^{i\hat{C}}$. Now define the quantity,

$$I = \left\langle \Omega \left| \hat{U}^\dagger \hat{F} \hat{U} \right| \Omega \right\rangle \qquad (2.7)$$

where $|\Omega\rangle$ is a yet unspecified state vector. Use (1.2) and (1.4) in the above to obtain,

$$I = \left\langle \Omega \left| \hat{U}^\dagger \hat{F} \hat{U} \right| \Omega \right\rangle = \sum_{n,m} \left\langle \Omega \left| \hat{U}^\dagger \right| \varphi_n \right\rangle \left\langle \varphi_n \left| \hat{F} \right| \varphi_m \right\rangle \left\langle \varphi_m \left| \hat{U} \right| \Omega \right\rangle = \sum_n F_n \left| \left\langle \varphi_m \left| \hat{U} \right| \Omega \right\rangle \right|^2 \tag{2.8}$$

Now if $F_n \geq 0$ for all $\left| \varphi_n \right\rangle$ then $I \geq 0$. However, if $I < 0$ then there must exist at least one $F_n$ which is negative. Therefore we will investigate whether or not $I$ can be negative. To evaluate (2.7) we will use the Baker-Campell-Hausdorff relationships which states that,

$$e^{+\hat{O}_1} \hat{O}_2 e^{-\hat{O}_1} = \hat{O}_2 + \left[ \hat{O}_1, \hat{O}_2 \right] + \frac{1}{2} \left[ \hat{O}_1, \left[ \hat{O}_1, \hat{O}_2 \right] \right] + \dots \tag{2.9}$$

Use this to obtain,

$$I = \left\langle \Omega \left| e^{i\hat{C}} \hat{F} e^{-i\hat{C}} \right| \Omega \right\rangle = \left\langle \Omega \left| \left( \hat{F} + \left[ i\hat{C}, \hat{F} \right] + \frac{1}{2} \left[ i\hat{C}, \left[ i\hat{C}, \hat{F} \right] \right] + \dots \right) \right| \Omega \right\rangle \tag{2.10}$$

In order to evaluate $\left[ i\hat{C}, \hat{F} \right]$ use (1.3), (2.1), and (2.3) to obtain,

$$\left[ \hat{F}, \nabla \times \hat{\mathbf{E}}_T \right] = -i \left( \nabla \times \nabla \times \hat{\mathbf{B}} - \nabla \times \hat{\mathbf{j}} + \frac{\partial \nabla \times \hat{\mathbf{E}}_T}{\partial x_1} \right) \tag{2.11}$$

and

$$\left[ \hat{F}, \hat{\mathbf{B}} \right] = i \left( \nabla \times \hat{\mathbf{E}}_T - \frac{\partial \hat{\mathbf{B}}}{\partial x_1} \right) \tag{2.12}$$

From these relationships and (2.6) we obtain,

$$\left[ \hat{F}, \hat{C} \right] = i \int \left[ \begin{array}{l} \left( \mathbf{D} \cdot \left( -\nabla \times \nabla \times \hat{\mathbf{B}} + \nabla \times \hat{\mathbf{j}} - \dfrac{\partial \nabla \times \hat{\mathbf{E}}_T}{\partial x_1} \right) \right) \\ + \mathbf{L} \cdot \left( \nabla \times \hat{\mathbf{E}}_T - \dfrac{\partial \hat{\mathbf{B}}(\mathbf{x})}{\partial x_1} \right) \end{array} \right] d^3 x \tag{2.13}$$

Rearrange terms and integrate by parts to obtain,

$$\left[ \hat{F}, \hat{C} \right] = i \int \left\{ \begin{array}{l} \left( \mathbf{D} \cdot \nabla \times \hat{\mathbf{j}} \right) - \hat{\mathbf{B}} \cdot \left[ \left( \nabla \times \nabla \times \mathbf{D} \right) - \dfrac{\partial \mathbf{L}}{\partial x_1} \right] \\ + \hat{\mathbf{E}}_T \cdot \left[ \nabla \times \mathbf{L} + \dfrac{\partial \nabla \times \mathbf{D}}{\partial x_1} \right] \end{array} \right\} d^3 x \tag{2.14}$$

Assume we can find a $\mathbf{D}(\mathbf{x})$ and $\mathbf{L}(\mathbf{x})$ such that,

$$\left(\nabla\times\nabla\times\mathbf{D}\right)-\frac{\partial\mathbf{L}}{\partial x_1}=0 \tag{2.15}$$

and

$$\nabla\times\mathbf{L}+\frac{\partial\nabla\times\mathbf{D}}{\partial x_1}=0 \tag{2.16}$$

These two equations are satisfied by,

$$\mathbf{D}(\mathbf{x})=\mathbf{e}_3 fW(x_1);\qquad \mathbf{L}(\mathbf{x})=-\mathbf{e}_3 f\,\partial_1 W(x_1) \tag{2.17}$$

where $\mathbf{e}_j$ is a unit vector in the $x_j$ direction, $W(x_1)$ is an arbitrary real valued function, and $f$ is a constant. Use these results in (2.14) to obtain,

$$\left[i\hat{C},\hat{F}\right]=\int\left(\mathbf{D}\cdot\nabla\times\hat{\mathbf{j}}\right)d^3x=\int\left(\hat{\mathbf{j}}\cdot\nabla\times\mathbf{D}\right)d^3x=-f\int\left(\partial_1 W(x_1)\hat{\mathbf{j}}\cdot\mathbf{e}_2\right)d^3x \tag{2.18}$$

where we have used $\nabla\times\mathbf{D}=-\mathbf{e}_2 f\,\partial_1 W(x_1)$. Next use the fact that the gauge fields $\hat{\mathbf{E}}_\perp$ and $\hat{\mathbf{B}}$ commute with the fermion fields $\hat{\psi}$ and $\psi^\dagger$ [4] so that

$$\left[\hat{\mathbf{E}}_\perp(\mathbf{x})\text{ or }\hat{\mathbf{B}}(\mathbf{x}),\hat{\mathbf{j}}(\mathbf{y})\right]=0.$$ Use this to obtain $\left[i\hat{C},\left[i\hat{C},\hat{F}\right]\right]=0$. Therefore,

$$I=\left\langle\Omega\left|e^{i\hat{C}}\hat{F}e^{-i\hat{C}}\right|\Omega\right\rangle=\left\langle\Omega\left|\hat{F}\right|\Omega\right\rangle-f\int\left(\left\langle\Omega\left|\hat{\mathbf{j}}\right|\Omega\right\rangle\cdot\mathbf{e}_2\partial_1 W(x_1)\right)d^3x \tag{2.19}$$

Integrate by parts to obtain,

$$I=\left\langle\Omega\left|\hat{F}\right|\Omega\right\rangle+f\int W(x_1)\partial_1 j_{2,e}(\mathbf{x})d^3x \tag{2.20}$$

where $j_{2,e}=\left\langle\Omega\left|\hat{\mathbf{j}}\right|\Omega\right\rangle\cdot\mathbf{e}_2$. The quantity $j_{2,e}$ is the expectation value of the $x_2$ component of current.

Now so far $|\Omega\rangle$ is an arbitrary state and $W(x_1)$ is an arbitrary function. Let us assume we can pick $|\Omega\rangle$ and $W(x_1)$ so that,

$$\int W(x_1)\partial_1 j_{2,e}(\mathbf{x})d^3x\neq 0 \tag{2.21}$$

How do we know that we can find a $W(x_1)$ and $|\Omega\rangle$ so that the above condition is satisfied? If QED is a correct model of the real world than states must exist for which the above relationship holds because the above relationship can be true in classical physics which is often a very close approximation to the real world.

Now examine Eq. (2.20). Note that $\langle \Omega | \hat{F} | \Omega \rangle$ and the integrated term $\int W(x_1) \partial_1 j_{2,e}(\mathbf{x}) d^3 x$ are independent of $f$ so that we can change $f$ without changing these other terms. Therefore, since the integrated term is nonzero, it is evident that we can always find an $f$ so that the quantity $I$ is negative. For instance if the integrated term is negative(positive) and if $f$ is a positive (negative) number with a large enough magnitude then $I$ will be negative. This means that there must exist at least one eigenstate $|\varphi_n\rangle$ where $F_n < 0$. For this eigenstate, then, the energy-momentum 4-vector will be space-like. This is consistent with the results of Ref. [4].

## 3. Discussion.

The existence of states with space-like energy momentum strongly implies the existence of states with negative energy with respect to the vacuum state. This is because it is always possible to find a Lorentz transformation that will convert a space-like state into a negative energy state. This contradicts the generally held belief that the vacuum state is the state of minimum energy. If this belief is true then this suggests that there is an error in the results of this paper. Therefore we must consider the status of this belief.

Now how would one prove that the vacuum state for a Dirac-Maxwell field in the coulomb gauge is the minimum energy state? The most direct way would be to solve the equation,

$$\hat{H} |\varphi_n\rangle = E_n |\varphi_n\rangle \tag{3.1}$$

where $\hat{H}$ is the Hamiltonian for the Dirac-Maxwell field which is given in [4] as,

$$\hat{H} = \int \left\{ -i\hat{\psi}^\dagger \boldsymbol{\alpha} \cdot \nabla \hat{\psi} + m\hat{\psi}^\dagger \beta \hat{\psi} \right\} d\mathbf{x} - \int \hat{\mathbf{j}} \cdot \hat{\mathbf{A}} d\mathbf{x} + \frac{1}{2} \int d\mathbf{x} \int d\mathbf{x}' \frac{\hat{\rho}(\mathbf{x})\hat{\rho}(\mathbf{x}')}{4\pi |\mathbf{x} - \mathbf{x}'|} + \frac{1}{2} \int \left( \hat{\mathbf{E}}_\perp^2 + \hat{\mathbf{B}}^2 \right) d\mathbf{x} + \varepsilon_R \tag{3.2}$$

where $\hat{\rho} = \hat{\psi}^\dagger \hat{\psi}$ is the charge operator, $\hat{\mathbf{A}}$ is the operator for the vector potential, and $\varepsilon_R$ is a renormalization constant so that the energy of the vacuum state is zero.

If we could solve equation (3.1) and obtain the eigenstates $|\varphi_n\rangle$ and eigenvalues $E_n$ then we could then identify which of the eigenstates $|\varphi_n\rangle$ was the vacuum state and determine if its eigenvalue was less than all the other eigenvalues. The problem is that, given the complexity of the Hamiltonian in (3.2), equation (3.1) is very complicated and can't be solved. In fact, as far as I can tell, there exists no proof that the vacuum state is the minimum energy state for a Dirac-Maxwell field in the coulomb gauge. The widely held belief that the vacuum state is the minimum energy state is merely a *conjecture* with no mathematical foundation.

In fact recent work has shown that the conjecture that the vacuum state is the state of minimum energy in quantum mechanics may not be correct for certain situations. For example, it has recently been shown [6] that the widely held belief that the vacuum state is the minimum energy state in Dirac's hole theory is not correct. It was shown in [6] that there must exist states with less energy than the vacuum in hole theory. Also it was shown in [7] that for a Dirac-Maxwell field in the temporal gauge there exsist states with less energy than the vacuum state.

Therefore the conjecture that the vacuum state is the minimum energy state is not grounds for rejecting the mathematical results derived in this paper and that of Ref. [4] because this conjecture is merely an assumption without mathematical proof. In fact the results of this paper (and that of [4]) strongly suggest that this conjecture is not correct for a Dirac-Maxwell field in the coulomb gauge.

**References**

1. R. Haag, *Local Quantum Physics*, Springer-Velag, Berlin (1993).

2. K. Nishijima, *Field and Particles:Field theory and Dispersion Relationships*, W.A. Benjamin, New York, 1969.

3. S. Weinberg, *The Quantum Theory of Fields Vol. 1*, Press Syndicate of the University of Cambridge, Cambridge, (1995).

4. D. Solomon, CEJP, **4**(3) 2006, p380-392. See also arXiv/hep-th/0412137.

5. E. V. Stefanovich, *Relativistic Quantum Dynamics: A non-traditional perspective on space, time, particles, fields, and action-at-a-distance.* arXiv/physics/0504062. (see Section 3.2.4).

6. D. Solomon. Physc. Scr. **74** (2006) 117-122. (see also quant-ph/0607037)

7. D. Solomon. Apeiron, **13,** No. 2, 240 (2006). (see also hep-th/0404193)